\documentclass[prx,showpacs,superscriptaddress,floatfix,twocolumn]{revtex4-2}

\usepackage{epsfig}
\usepackage{amsmath}
\usepackage{amssymb}
\usepackage{amsfonts}
\usepackage{mathptmx}
\usepackage{eucal}
\usepackage{bm}
\usepackage{color}
\usepackage[colorlinks,linkcolor=blue,citecolor=blue]{hyperref}
\usepackage{epstopdf}
\usepackage{bbold}
\usepackage{url}
\usepackage{svg}

\def\eps{\varepsilon}
\def\up{\uparrow}
\def\dw{\downarrow}
\def\bk{{\mathbf{k}}}
\def\br{{\mathbf{r}}}

\begin{document}

\title{
High orbital-moment Cooper pairs by crystalline symmetry breaking}

\author{Maria Teresa Mercaldo}
\affiliation{Dipartimento di Fisica ``E. R. Caianiello", Universit\`a di Salerno, IT-84084 Fisciano (SA), Italy}

\author{Carmine Ortix}
\affiliation{Dipartimento di Fisica ``E. R. Caianiello", Universit\`a di Salerno, IT-84084 Fisciano (SA), Italy}

\author{Mario Cuoco}
\affiliation{SPIN-CNR, IT-84084 Fisciano (SA), Italy, c/o Universit\`a di Salerno, IT-84084 Fisciano (SA), Italy}

\begin{abstract}
The pairing structure of superconducting materials is regulated by the point group symmetries of the crystal. 
Here, we study spin-singlet multiorbital superconductivity in materials with unusually low crystalline symmetry content
and unveil the the appearance of  even-parity (s-wave) Cooper pairs with high orbital moment. 
We show that the lack of mirror and rotation symmetries makes
pairing states with quintet orbital angular momentum symmetry-allowed.
A remarkable fingerprint of this type of pairing state is provided by a nontrivial superconducting phase texture in momentum space with $\pi$-shifted domains and walls with anomalous phase winding. The pattern of the quintet pairing texture is shown to 
depend on the orientation of the orbital polarization and the strength of the mirror and/or rotation symmetry breaking terms.
Such momentum dependent phase makes Cooper pairs with net orbital component 
suited to design orbitronic Josephson effects. We discuss how an intrinsic orbital dependent phase can set out anomalous Josephson couplings by employing superconducting leads with nonequivalent breaking of crystalline symmetry.  
\end{abstract}

\maketitle

\section{Introduction}
\begin{figure*}[t!]
\begin{center}
\includegraphics[width=0.9\textwidth]{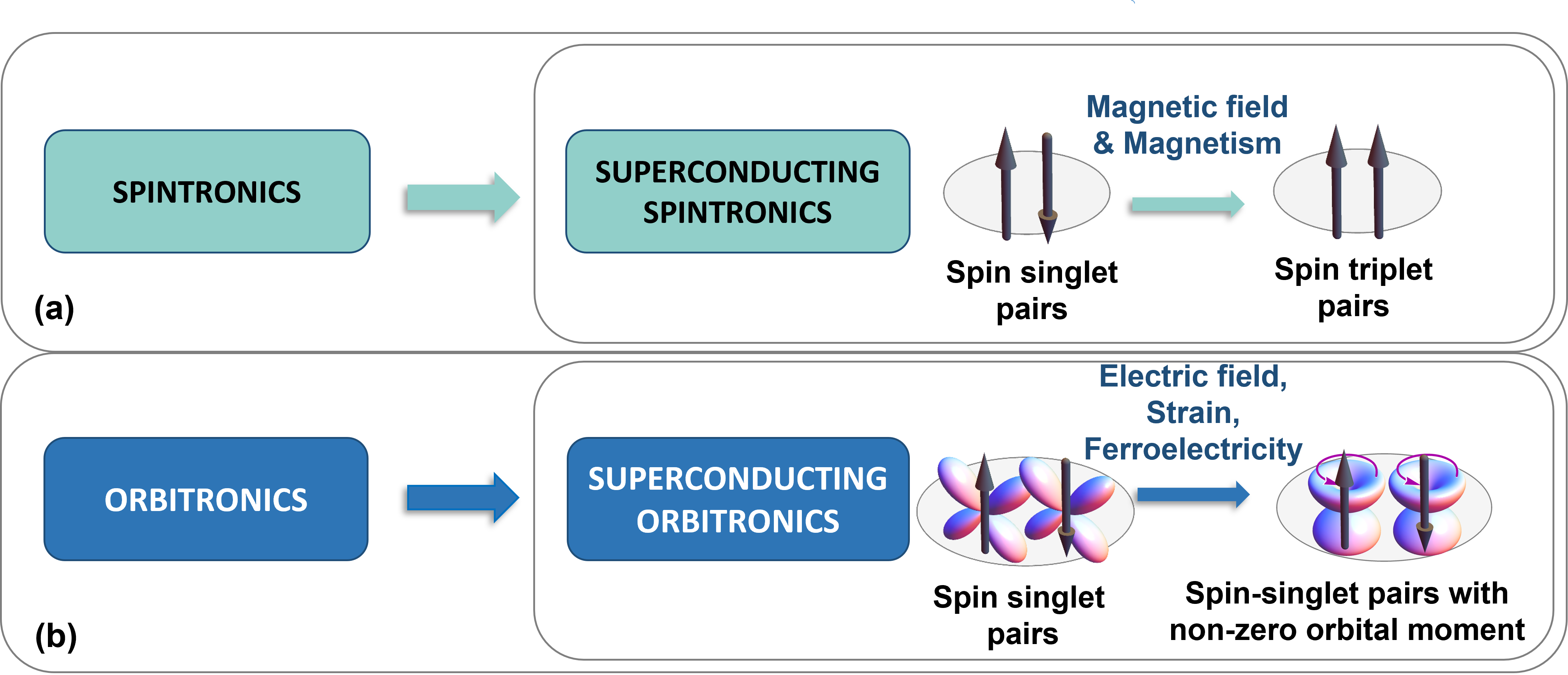} 
\protect\caption{Schematic representation of the analogy between superconducting spintronics and superconducting orbitronics, highlighting the physical resources for achieving spin polarized and orbitally polarized supercurrents. (a) Spintronics exploits the intrinsic spin of the electron and its associated magnetic moment, rather than the electron charge, for transport and storage of information \cite{Zutic2004,Fert2008}. The spintronic circuits employ spin currents to design logic operations and devices that perform faster and with greater energy efficiency than the charge-based equivalent in semiconductor technologies. A synergy of superconductivity and spintronics in superconducting spintronic devices is based on the successful conversion of spin-singlet into spin-triplet Cooper pairs, e.g. at the superconductor-magnet interface or in general through magnetic sources. As spin-triplet Cooper pairs are spin-polarized, it follows that the resulting supercurrents can carry a net spin component without the harmful heating effects of spintronic devices. (b) Orbitronics represents a next generation device technology which exploits the orbital content of the electronic states and in particular the orbital current as an information carrier \cite{Go_2021}. An orbital current is made by the flow of electrons with a finite orbital angular momentum. Being related to the orbital states, the orbital current is prone to be electrically generated and transported. In superconducting orbitronics orbital supercurrents can carry a net component of the orbital angular momentum. Differently from the spin-poralized supercurrents that make use of magnetic ordering and magnetic fields, orbital supercurrents can be obtained by means of electrical or mechanical means as they generate interactions that reduce the point group crystalline symmetry.}
\label{fig:1}
\end{center}
\end{figure*}

Cooper pairs in superconducting materials have been originally introduced in two possible existing configurations: spin-singlet and spin-triplet \cite{Cooper_1956}.
The appearance of spin-triplet superconductivity \cite{Sigrist_1991} in pristine materials is however very rare.
For this reason, spin-polarized Cooper pairs have been mostly searched in 
hybrid heterostructures where conventional spin-singlet superconductors 
are interfaced with a magnetic material \cite{Bergeret2005, Linder2015,Eschrig_2015,Amundsen2023}.
Generation
and control of spin-polarized supercurrents 
is at the basis of
superconducting spintronics
 -- a form of electronics using dissipationless supercurrents and thus characterized by extreme low-energy consumption (Fig. \ref{fig:1}a). 
The intrinsic vectorial nature of the spin-triplet order parameter makes 
superconducting spintronic devices suitable for functionalization with external
Zeeman fields \cite{Sengupta2001,Romano2010,Brydon2010,Hyart2014,Mercaldo_PhysRevB.94.140503,Mercaldo2019a} or 
magnetic materials.
\cite{Nakosai_PhysRevB.88.180503,Brydon2011,Gentile2013,Mercaldo2018a,Mercaldo2018b,Maistrenko2021,Poniatowski2022}.
Spin-triplet pairing has also a strong link 
to topological superconductivity \cite{Read2000,Kitaev2003,Sato_2017,Sato2009,Qi_2011_RevModPhys.83.1057}. This is evident in the paradigmatic one-dimensional (1D) Kitaev model \cite{Kitaev2003}, which describes Cooper pairs with equal spin and can harbor Majorana bound states at the chain ends.
Majorana boundary modes are key entities for the design of topological quantum computation \cite{Ivanov2001,Kitaev2003,Nayak2008}. 
This has triggered great interest in 
topological superconductors \cite{Frolov2020,Leijnse_2012,Kotetes_2013} and 
more generally spin-triplet pairing. 

The physical description for the appearance of spin polarized pairs generally relies on an effective electronic structure involving a single Kramers' pair of bands.
Nevertheless, multiorbital electronic states 
can be present close to the Fermi level. 
In this case, Cooper pairs can carry  a total
angular momentum that is related to the combination of the spin and orbital moment rather than only spin.
In superconductors with strong spin-orbit coupling 
electrons at the Fermi level can have an effective angular momentum 
larger than $1/2$ \cite{Brydon_2016,Kim_2018}. This 
occurs in some cubic half-Heusler materials (e.g. RPtBi or RPdBi with R being the rare earth element), where the electrons at the Fermi level have
a total angular momentum $3/2$ due to high crystal symmetry and spin-orbit interaction \cite{Brydon_2016,Kim_2018}. The effective spin $3/2$ character of the 
electrons participating in the Cooper pairs can yield exotic spin quintet (J=2) and spin septet (J=3) \cite{Brydon_2016,Kim_2018,Venderbos_2018,Yu_2018,Roy_2019} configurations 
besides the more conventional singlet and triplet pairing states.

In this study, we
take a different route and consider a scenario where high orbital-moment Cooper pairs occur in a regime of vanishing spin-orbit coupling
and are, instead, induced by the breaking of mirror and rotation symmetries. 
Understanding the mechanisms for the generation and control of Cooper pairs with net components of the orbital angular momentum 
sets a milestone for the development 
of superconducting orbitronics (Fig. \ref{fig:1}b). 
Orbitronics is a new form of electronics that 
exploits the orbital content of the electronic states and 
orbital currents as information carriers \cite{Go_2021}.
In superconductors, these currents (and more generally the role of orbital degrees of freedom) have been discussed  considering spin-singlet pairing.
Orbital degrees of freedom are particularly relevant in acentric superconductors due to the presence of the so-called 
orbital Rashba coupling~\cite{park11,park12,Khalsa2013PRB,kim14,Mercaldo2020}. 
In two-dimensional (2D) or quasi-two-dimensional systems, the absence of inversion symmetry and consequently of an horizontal
mirror symmetry ($\mathcal{M}_z$) defines a polar axis ${\hat z}$ 
and yields an orbital Rashba coupling
($\alpha_\mathrm{OR}$) that mixes orbital configurations with different mirror parity.
Precisely as the well-known Rashba spin-orbit coupling,  the orbital Rashba 
term couples the atomic angular momentum ${\bf L}$ with the crystal wave-vector $\bf{k}$ ~\cite{park11,park12,Khalsa2013PRB,kim14,Mercaldo2020}.
In layered superconductors, the orbital Rashba coupling has been shown 
to drive an orbital reconstruction of the superconducting phase by inducing a so called 0-$\pi$ transition. The induced $\pi$-paired phase is marked by a non-trivial sign change of the superconducting order parameter between different orbitals \cite{Mercaldo2020}. Such orbital antiphase has physical consequences in
the tunneling \cite{Mercaldo2021_a}, the magnetic response \cite{Bours2020} and the Josephson effects \cite{Guarcello2022}. Moreover, 
in 2D spin-singlet superconductors with unusually low crystalline symmetry,
vortices with supercurrents carrying orbital angular momentum \cite{Mercaldo2022} around the core have been shown to be 
energetically stable. 
The orbital Rashba coupling has been also shown to trigger
an orbitally driven Edelstein effect \cite{Chirolli2022} 
resulting in an orbital 
polarization one order of magnitude larger than the spin polarization
\cite{Chirolli2022}. 
For acentric multiorbital superconductors with the spin-triplet pairing involving only different orbitals, an even-parity s-wave topological phase can be achieved \cite{fukaya18} with an intrinsic tendency to yield 0-, $\pi$-, and $\phi$- Josephson phase couplings together with dominant high-harmonics in the current phase relation \cite{fukaya19,fukaya20,fukaya22}.

Motivated by these advances in the study of
superconducting orbitronics effects, in this manuscript we demonstrate how in 2D supercondutors with low crystalline symmetries Cooper pairs with equal orbital moment can be formed. In particular, we
show that Cooper pairs with net orbital component and maximal orbital polarization can 
possess a pairing amplitude texture in momentum space due to the nontrivial phase imprinted by the combination of pairing components with equal and different orbital character. The texture of the pairing amplitude in momentum space can be controlled by suitably varying the strength of the coupling terms that 
break mirror or rotation symmetry. 
Such fingerprint allows to design Josephson orbitronics effects which exploit the tunneling of Cooper pairs with net orbital moment.

\section{Model and orbital dependent order parameters}

\begin{figure*}[hbt]
\begin{center}
\includegraphics[width=0.97\textwidth]{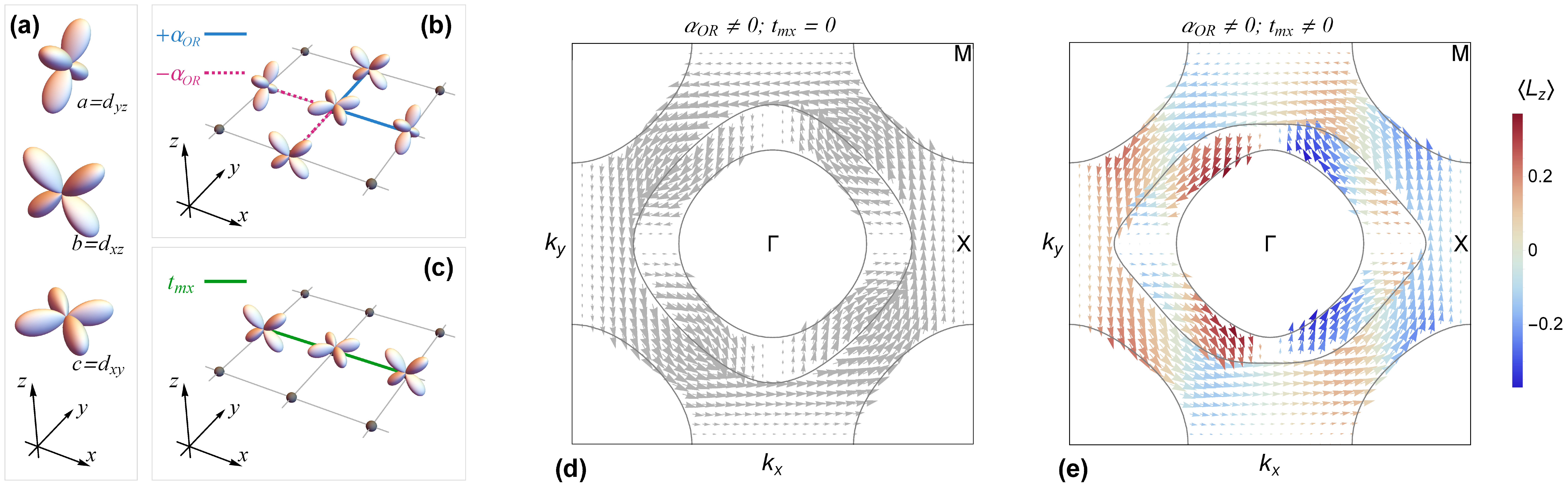}
\protect\caption{(a) Schematic of the orbital states considered in the multiorbital model. (b) Sketch of the orbital mixing hopping on the lattice associated with the orbital Rashba coupling. Due to the ($\mathcal{M}_z$) horizontal mirror symmetry breaking the orbital hybridization is between states with unequal $\mathcal{M}_z$ mirror parity and it is odd in space for exchange of the $(x,y)$ coordinates. (c) Scheme of the orbital hybridization related to the term $t_{mx}$ that breaks all the mirror symmetries and rotations except $\mathcal{M}_x$. (d) Texture of the orbital angular momentum in the Brillouin zone for the electronic structure in the normal state. The lines correspond to the Fermi contour at a representative chemical potential $\mu=-0.4t$. The pattern corresponds to have only in plane components of the orbital moment, i.e. $L_x$ and $L_y$, as the horizontal mirror symmetry and the inversion symmetry are broken due to the nonvanishing obital Rashba interaction, namely $\alpha_{OR}=1.0t$. (e) Texture of the orbital angular momentum in the Brillouin zone for a representative value of $t_{mx}=0.4 t$. Due to the vertical mirror and rotation symmetry breaking, an out-of-plane value of the orbital angular momentum is obtained that has a d-wave like pattern occurring along the diagonals of the Brillouin zone.}
\label{fig:2}
\end{center}
\end{figure*}

We consider a multi-orbital model with electronic states
that span an $L=1$ manifold. For convenience, we choose Wannier functions
that can describe $d-$ or $p-$ orbitals which we indicate as ($a,b,c$) and span 
the manifold for each spin configuration.
The orbitals are localized at the site of a square lattice (Fig. \ref{fig:2}) with $\mathcal{C}_{4v}$ point group symmetry. We work in the basis of Wannier states of zero angular momentum configurations
$|0\rangle_{l}$  ($l=(a,b,c)$),
 i.e. ${_\alpha}\langle 0|\hat{L}| 0\rangle_\alpha =0$. 
In this basis, the angular momentum components are expressed as $\hat L_k=i \eps_{klm}$,
with $\eps_{klm}$ being the Levi-Civita tensor. The absence of horizontal mirror symmetry
$\mathcal{M}_z$
and hence of inversion due to, 
{\it e.g.} 
structural inversion asymmetry or
external electric fields
yields an orbital Rashba term
that couples the atomic angular momentum ${\bf L}$ with the crystal wave-vector ${\bf k}$. We also assume that the
vertical mirror symmetries ($\mathcal{M}_x$ and $\mathcal{M}_y$) as well as the $C_4$ rotation around the $\hat{z}$ axis are
broken for instance by local built-in elecric fields or (inhomogeneous) strains. This additional symmetry lowering
can be included by terms that involve the product of 
different orbital angular momentum components. By further
assuming a conventional s-wave spin-singlet pairing due to a local attraction, the Hamiltonian in momentum space can be expressed as \cite{Mercaldo2020,Bours2020,Mercaldo2021_a}
\begin{eqnarray}
\mathcal{H}= \sum_{\bk} \Psi^{\dagger}(\bk) \hat{H}(\bk) \Psi(\bk) \,,
\label{ham}
\end{eqnarray}
where $\hat{H}(\bk)= (\hat{T}_{k} - \mu ) \tau_z + {\hat{\Delta} \tau_+ +\hat{\Delta}^\dagger\tau_-} $
within the spinorial basis $\Psi^\dagger(\bk) =(\Psi_{\up}^\dagger(\bk), \Psi_{\dw}(-\bk))$ assuming that they are related to the orbital basis as 
$\Psi_{\sigma}^\dagger(k)  =
(c^\dagger_{a\sigma}(\bk),c^\dagger_{b\sigma}(\bk),c^\dagger_{c\sigma}(\bk))$,  $\tau_i$ are the Pauli matrices in the particle-hole space, {with $\tau_\pm=(\tau_x\pm i \tau_y)/2$}. Here, 
$c_{\alpha\sigma}(\bk)$
is the canonical electron annihilation operator with momentum $\bk$,  spin $\sigma$ and zero orbital polarization components, with $\alpha=(a,b,c)$ {e.g. $=(d_{yz},d_{xz},d_{xy})$ (Fig. \ref{fig:2}a}).
The kinetic energy matrix $\hat{T}_{k}$ is expressed as 
\begin{eqnarray}
{\hat{T}_{k}}&=&\sum_{j=x,y,z}  \epsilon_j({\bf k})\mathbb{P}_j - \alpha_{\text{OR}} (\sin(k_x) \hat{L}_y - \sin(k_y) \hat{L}_x) \nonumber \\ 
&-&2t_{mx}\cos(k_x)  \{\hat{L}_y,\hat{L}_z\} -2t_{my} \cos(k_y) \{\hat{L}_x,\hat{L}_z\} \;,
\end{eqnarray} 
where $\epsilon_x({\bf k})=-2t(\gamma \cos(k_x)+\cos(k_y))$, 
$\epsilon_y({\bf k})=-2t( \cos(k_x)+\gamma\cos(k_y))$, and $\epsilon_z({\bf k})=-2t(\cos(k_x)+\cos(k_y))$ are dispersion relations,
$\mathbb{P}_j=(\hat{L}^2/2-\hat{L}_j^2)$ and 
 $\gamma\in[0,1]$  
is a coefficient of anisotropic terms which are compatible with a $\mathcal{C}_{4v}$ point group.
$\alpha_{\text{OR}}$ is the strength of the orbital Rashba coupling, $\{\hat{A},\hat{B}\}$  denotes  the anticommutator, and $t_{mx(my)}$ is an additional inter-orbital nearest-neighbor hopping.
As we discuss below, $\alpha_{\text{OR}}$ and $t_{mx(my)}$ have a different symmetry content.
Taking into account the expressions in momentum space for the terms that break mirror and rotation symmetries it is useful to explicitly indicate the electronic processes in real space and their orbital dependence.
In order to do so, we introduce a basis at a given site ${\bf{r}}_i$ of the lattice as $(c_{a,\sigma,i},c_{b,\sigma,i},c_{c,\sigma,i})$ (with $c_{\alpha,\sigma,i}\equiv c_{\alpha,\sigma}(\br_i)$).
Then, within this representation the single particle terms in the Hamiltonian  can be separated in two parts (${\cal H}_{x}$ and ${\cal H}_{y}$) for 
the $x$ and $y$ directions, respectively. Assuming two nearest neighbor centers at positions $\br_i$ and $\br_j$, the real space Hamiltonian is given by:
\begin{eqnarray}
    {\cal H}_{x}(\br_i,\br_j) &=& \sum_{\sigma=\up,\dw}[-2t( \gamma c^\dagger_{a\sigma,i} c_{a\sigma,j} + c^\dagger_{b\sigma,i} c_{b\sigma, j}  +  c^\dagger_{c\sigma,i} c_{c\sigma,j})  \nonumber  \\ \nonumber
    &-&\alpha_{OR} (c^\dagger_{a\sigma,i} c_{c\sigma,j} - c^\dagger_{c\sigma,i} c_{a\sigma,j}) \\ 
    &-& t_{mx} (c^\dagger_{b\sigma,i} c_{c\sigma,j} + c^\dagger_{c\sigma,i} c_{b\sigma,j}) ] + H.c.
\end{eqnarray}
for sites along the $x$ direction, with $\br_i=(r_{i_x},r_{i_y})$ and $\br_j=(r_{i_x+1},r_{i_y})$.
In analogous way one can write down the real space Hamiltonian along the $y$ direction, 
\begin{eqnarray}
    {\cal H}_{y}(\br_i,\br_j) &=& \sum_{\sigma=\up,\dw}[-2t( c^\dagger_{a\sigma,i} c_{a\sigma,j} + \gamma c^\dagger_{b\sigma,i} c_{b\sigma, j}  +  c^\dagger_{c\sigma,i} c_{c\sigma,j})  \nonumber \\ \nonumber
    &-&\alpha_{OR} (c^\dagger_{b\sigma,i} c_{c\sigma,j} - c^\dagger_{c\sigma,i} c_{b\sigma,j}) \\ 
    &-& t_{my} (c^\dagger_{a\sigma,i} c_{c\sigma,j} + c^\dagger_{c\sigma,i} c_{a\sigma,j}) ] + H.c.
\end{eqnarray}
with $\br_i=(r_{i_x},r_{i_y})$ and $\br_j=(r_{i_x},r_{i_y+1})$ (see also Fig. \ref{fig:2}(b-c)). 
We point out that the electronic processes associated with $\alpha_{\text{OR}}$ and $t_{mx(my)}$ have a different orbital structure. For instance, $\alpha_{\text{OR}}$ yields a mixing of $xy$ and $yz$ ($xz$) orbitals on nearest neighbor sites along the $x$ ($y$) direction, respectively. Such hybridization process has additionally an odd spatial parity, i.e. it changes sign by exchanging the sites position. Instead, the crystalline potentials related to $t_{mx(my)}$ terms have an even spatial parity and are based on orbital hybridization among the $xy$ and $xz$ ($yz$) orbitals along the $x$ ($y$) directions, respectively.

Let us now specify the symmetry properties of the various terms entering into the kinetic energy. To do so we notice that upon a $C_4$ rotation $k_x \rightarrow k_y$, $k_y \rightarrow -k_x$ and the same applies to the angular momentum components, i.e. $\hat{L}_x \rightarrow \hat{L}_y$, $\hat{L}_y \rightarrow -\hat{L}_x$, $\hat{L}_z \rightarrow \hat{L}_z$. Furthermore, for the mirror symmetry $\mathcal{M}_y$ we have that $k_y \rightarrow -k_y$ ($k_x$ and $k_z$ stay unchanged) while the angular momentum is a pseudovector and hence the component perpendicular to the mirror plane does not transform, {\it i.e.} $\hat{L}_y \rightarrow \hat{L}_y$, while the other are inverted, i.e. $\hat{L}_x \rightarrow -\hat{L}_x$ and $\hat{L}_z \rightarrow -\hat{L}_z$.
On the basis of these symmetry transformations, we have that $\alpha_{\text{OR}}$ breaks only the horizontal mirror symmetry $\mathcal{M}_z$ and mixes states with different $\mathcal{M}_z$ mirror parity (Fig. \ref{fig:2}b).
On the other hand $t_m$ terms break the vertical mirrors and rotation symmetries and involve the orbital mixing of states with different $\mathcal{M}_{x,y}$ mirror parity along the $x$ and $y$ directions (Fig. \ref{fig:2}c). 
In particular, if we consider only one of the two $t_{m}$ terms, there is a residual vertical mirror symmetry. In this case, 
the $\mathcal{C}_{4v}$ point group is reduced to $\mathcal{C}_{s}$. The presence of both $t_{mx}$ and $t_{my}$ breaks all point group symmetries except the identity: the point group is reduced to ${\mathcal C}_1$. 
Below, we analyze firstly the configurations which are $\mathcal{C}_{4v}$ and then consider the symmetry breaking terms to have the changeover from $\mathcal{C}_{4v}$ to $\mathcal{C}_{s}$.

Concerning the superconducting order parameter, 
we consider a conventional scenario with only local s-wave spin-singlet pairs with intra- and inter-orbital components. 
We assume that the superconducting phase is due to a local attraction that induces a spin-singlet s-wave pairing with intra- and inter-orbital character. 
The corresponding attractive local interaction has the following form:
\begin{equation}
H_{g}=-\sum_{i;\alpha \beta} g_{\alpha\beta} \, n_{\alpha}(i)\,n_{\beta}(i).
\end{equation}
We subsequently perform a 
conventional mean-field decoupling by introducing the superconducting pairing amplitudes in the spin-singlet channel. 
Then, we have that 
\begin{equation}
H_{g}=-\sum_{\bk;\alpha \beta} g_{\alpha\beta} \, f_{\alpha\beta}(\bk) c^{\dagger}_{\alpha \dw}(\bk) c^{\dagger}_{\beta \up}(-\bk)
\end{equation}
with
$f_{\alpha \beta}(\bk)=\langle c_{\alpha \up}(\bk) c_{\beta \dw}(-\bk) \rangle$ being the k-dependent orbital resolved pairing amplitude 
and
$\Delta_{\alpha \beta}=\int_{BZ}\frac{d^2k}{(2\pi)^2} g_{\alpha\beta} f_{\alpha \beta}(\bk)$ the corresponding $s$-wave order parameters.
The matrix $\hat{\Delta}$ has the following structure:
\begin{eqnarray}
    \hat{\Delta}=\left(
\begin{array}{c c c}
    \Delta^{}_{aa} & \Delta^{}_{ab} & \Delta^{}_{ac} \\
    \Delta^{}_{ba} & \Delta^{}_{bb} & \Delta^{}_{bc}  \\ \Delta^{}_{ca} & \Delta^{}_{cb} & \Delta^{}_{cc}
\end{array}
\right)\;.
\label{delta_mat}
\end{eqnarray}

\begin{figure*}[t!]
\begin{center}
\includegraphics[width=0.98\textwidth]{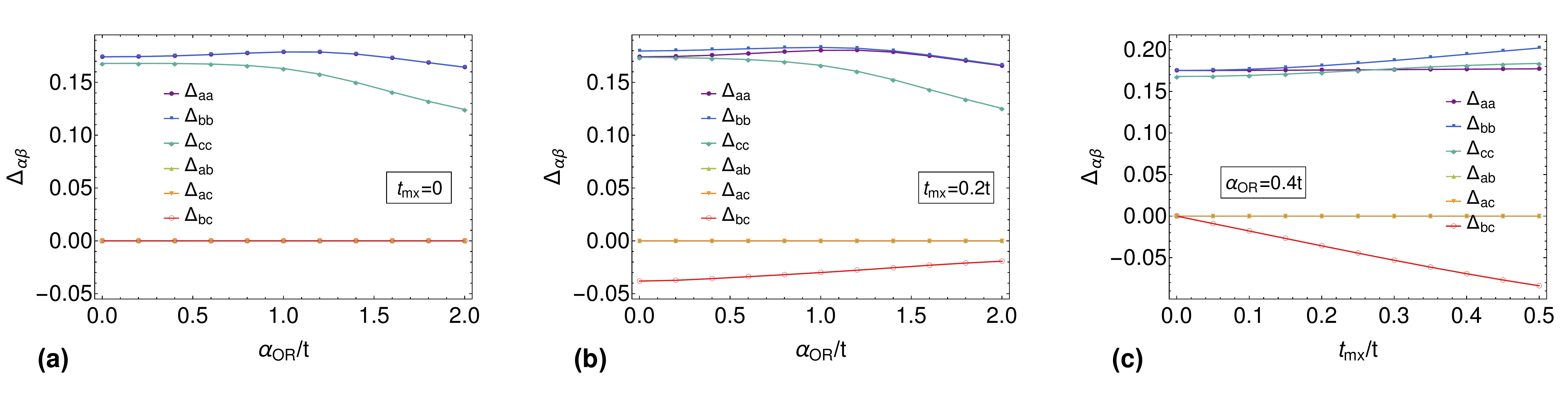} 
\protect\caption{Plots of the order parameter $\Delta_{\alpha\beta}$, obtained within a self-consistent analysis, as a function of the orbital Rashba coupling for (a) $t_{mx}=0$ and (b) for $t_{mx}=0.2t$. In (c) we show the evolution of the order parameters $\Delta_{\alpha\beta}$ as a function of the $t_{mx}$ amplitude at a given orbital Rashba coupling, namely $\alpha_{OR}=0.4t$. For the selected choice of the parameters we have that $\Delta_{ab}=\Delta_{ac}=0$, while  $t_{my}\neq 0$ would give $\Delta_{ac}\neq 0$. 
The SC-OPs $\Delta_{\alpha\beta}$ are real and symmetric  $\Delta_{\alpha\beta}= \Delta_{\beta\alpha}$. The other parameters are: $\mu=-0.4t$, $t_{my}=0$ and $\gamma=0.1$. }
\label{fig:3}
\end{center}
\end{figure*}

Here, $\langle ... \rangle$ stands for the zero temperature ground-state expectation value. Due to the structure of $\hat{\Delta}$ we observe that the intra- and inter-orbital components of the order parameters transform differently under the point group symmetries.
The intra-orbital terms $\Delta_{\alpha \alpha}$ 
are generally non-vanishing 
 -- the crystalline symmetries only pose constraints on their relative amplitudes.
On the other hand, the s-wave $\Delta_{\alpha\beta}$ with $\alpha \neq \beta$ transform as $(\hat{L}_{\alpha}\cdot \hat{L}_{\beta}+\hat{L}_{\beta}\cdot \hat{L}_{\alpha})$. 
Therefore they can pick up a $\pi$ phase upon rotation and vertical mirror symmetry. This immediately implies their vanishing values.
Rotation and mirror symmetry breaking 
is thus required to have a finite
$\Delta_{\alpha\beta}$ amplitude.
These aspects are demonstrated by performing a self-consistent computation of the orbital dependent order parameters for the $\mathcal{C}_{4v}$ and $\mathcal{C}_{s}$ point group symmetry as a function of $\alpha_{OR}$ and $t_{mx}$. The main outcomes are reported in Fig. \ref{fig:3}. We see that for the $\mathcal{C}_{4v}$ symmetry configuration only the diagonal components are nonvanishing and the orbital Rashba does not affect much the amplitude of the order parameter.
The lowering of the point-group symmetry
from $\mathcal{C}_{4v}$ to $\mathcal{C}_{s}$ due to the finiteness of 
the $t_{mx}$ coupling activates one off-diagonal component ($\Delta_{bc}$) whose amplitude is augmented as $t_{mx}$ increases.

\section{Cooper pairs with high-orbital moment}
\label{sec:s3}

The pairing operators in the basis of the quenched orbital angular momentum 
can be related to those with a finite net component of the orbital angular moment, 
which correspond to orbital quintets or orbital triplets.
Since we are assuming
spin-singlet pairing, the total spin moment is zero. Then, when considering the two-particle configurations one can restrict the analysis to the subspace spanned by the following product states $(a_{\sigma},b_{\sigma},c_{\sigma})\otimes(a_{\sigma'},b_{\sigma'},c_{\sigma'})$ with $\sigma=\uparrow (\downarrow)$ and $\sigma\neq \sigma'$. In this subspace, by combining the $L=1$ angular momenta in the $\uparrow$ and $\downarrow$ sectors we can construct paired states with total orbital moment $J$, with 
$J=0,1,2$. 
Taking into account the algebra for the sum of the angular momenta, one can introduce the operators for the pairing states
with given $(J,J_z)$ quantum numbers in terms of those associated with the quenched states of the type $\alpha_{\sigma}\,\beta_{\sigma'}$ with $\alpha$ and $\beta$ being $(a,b,c)$.
Indeed, we have a set of operators $p_{J,J_z}$ associated to paired states with a total orbital moment $J$ and a projection $J_z$ which are combination of those corresponding to paired states in the ($a,b,c$) basis. The same can be done for the maximal projection of $J_x$ and $J_y$. 
For convenience we report the explicit expression of the paired states with highest  $J_\alpha$ (i.e $J_\alpha=J$) for $J=1$ and 2. The other ones can be obtained by application of the lowering operator of the total orbital angular momentum operator. The explicit expressions for $p_{J,J_\alpha}$ are given by (see details for the construction of the pairing operators in the Appendix)
\begin{eqnarray}
\label{eq6}
p_{2,2_x}(\bk)&=&\frac{1}2 \{c_{c\up}(\bk)c_{c\dw}(-\bk)-c_{b\up}(\bk)c_{b\dw}(-\bk) \\ \nonumber 
&-&i[c_{b\up}(\bk)c_{c\dw}(-\bk)+c_{c\up}(\bk)c_{b\dw}(-\bk)]\}\\
p_{2,2_y}(\bk)&=&\frac{1}2 \{c_{c\up}(\bk)c_{c\dw}(-\bk)-c_{a\up}(\bk)c_{a\dw}(-\bk) \\ \nonumber 
&+&i[c_{a\up}(\bk)c_{c\dw}(-\bk)+c_{c\up}(\bk)c_{a\dw}(-\bk)]\}\\
p_{2,2_z}(\bk)&=&\frac{1}2 \{c_{b\up}(\bk)c_{b\dw}(-\bk)-c_{a\up}(\bk)c_{a\dw} (-\bk) \\ \nonumber 
&+&i[c_{a\up}(\bk)c_{b\dw}(-\bk)+c_{b\up}(\bk)c_{a\dw}(-\bk)]\} 
\label{eq:p22z}\\ \nonumber
\\
p_{1,1_x}(\bk)&=&\frac{1}2 [ c_{c\up}(\bk)c_{a\dw}(-\bk)-c_{a\up}(\bk)c_{c\dw}(-\bk) \\ \nonumber
&+&i (c_{a\up}(\bk)c_{b\dw}(-\bk)-c_{b\up}(\bk)c_{a\dw}(-\bk))]\\
p_{1,1_y}(\bk)&=&\frac{1}2 [ c_{c\up}(\bk)c_{b\dw}(-\bk)-c_{b\up}(\bk)c_{c\dw}(-\bk) \\ \nonumber
&+&i (c_{a\up}(\bk)c_{b\dw}(-\bk)-c_{b\up}(\bk)c_{a\dw}(-\bk))]\\
p_{1,1_z}(\bk)&=&\frac{1}2 [ c_{b\up}(\bk)c_{c\dw}(-\bk)-c_{c\up}(\bk)c_{b\dw}(-\bk) \\ \nonumber
&+&i (c_{a\up}(\bk)c_{c\dw}(-\bk)-c_{c\up}(\bk)c_{a\dw}(-\bk))]
\label{eq11}
\end{eqnarray}

Let us hence introduce the pairing amplitudes associated with the operators $p_{J,J_z}$ and indicate them as $f_{J,J_z}$ {(i.e. $f_{J,J_\alpha}=\langle p_{J,J_\alpha}\rangle$), $\alpha=(x,y,z)$}, similarly to the previously introduced operators of pairs with quenched orbital moment $c_{\alpha \sigma} c_{\beta \sigma'}$ {[i.e.  $f_{\alpha\beta}=\langle c_{\alpha \up}(\bk) c_{\beta \dw}(-\bk) \rangle$ 
 ]}.
Now, for the examined s-wave (even parity) spin-singlet configurations, we notice that the pairing amplitudes for the orbital triplet states, $(f_{1,1_\alpha},f_{1,0_\alpha},f_{1,-1_\alpha})$, are identically zero.
Indeed, for the spin-singlet state we have that, due to the antisymmetry of the fermionic correlator,  
$\langle c_{\alpha \up}(\bk) c_{\beta \dw}(-\bk) \rangle-\langle c_{\beta \up}(\bk) c_{\alpha \dw}(-\bk) \rangle=0$, with $\alpha\neq\beta$, $\alpha$ and $\beta$ being $(a,b,c)$. 
We observe that the orbital triplet pairing states with $J=1$ are symmetry allowed in the spin-triplet channel with a spatially odd-parity symmetry. This is because the pairing amplitudes has to be antisymmetric for exchange of the electrons.
Hence, the unique  non zero angular momentum s-wave spin-singlet states which are symmetry allowed are those with $J=2$. 
We also notice that both intra- and inter-orbital pairing amplitudes are crucial to give a nonvanishing value of the quintet state.
For instance, by inspection of the $f_{2,2_x}$ configuration, one needs to have that the orbital states which are mixed through $L_z$ should yield a pairing amplitude with $f_{bb}\neq f_{cc}$ or that $f_{bc}$ is non-vanishing.

\section{Orbital pairing textures in momentum space: lowering of point-group crystal symmetry}

\begin{figure*}[htb]
\begin{center}
\includegraphics[width=0.98\textwidth]{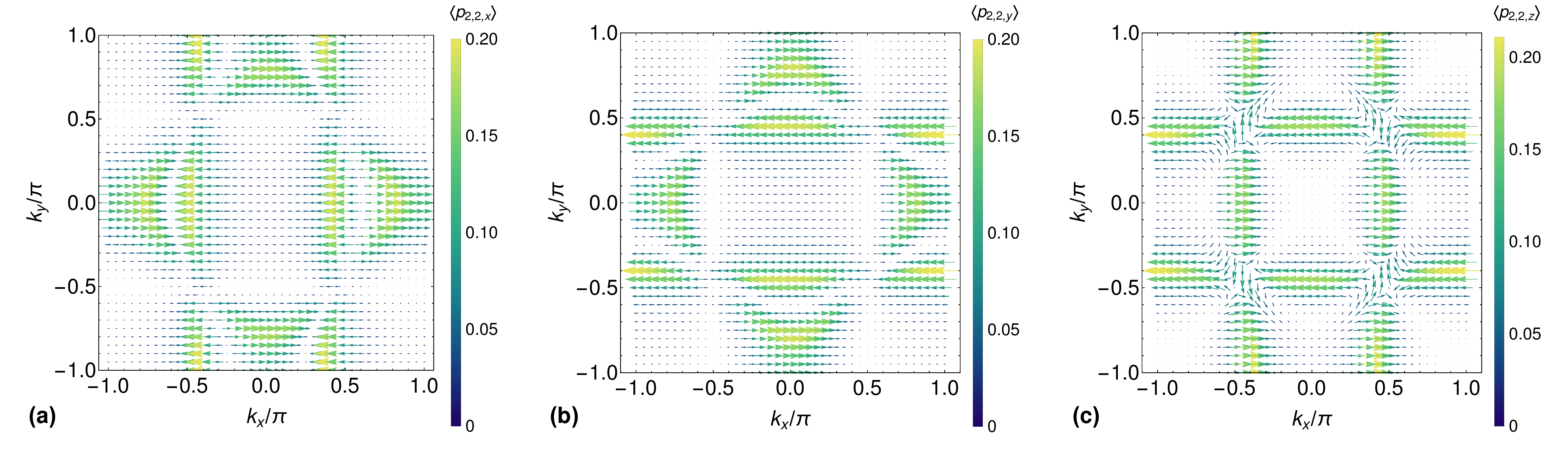} 
\protect\caption{(a)-(c) Vector plots of the pairing amplitudes $f_{2,2_\alpha}=\langle p_{2,2_\alpha}\rangle$ with ($\alpha=x,y,z$). 
The arrows represent  $(\rm{Re}\langle p_{2,2,\alpha}\rangle,\rm{Im}\langle p_{2,2,\alpha}\rangle)$. Here the parameters are  $\alpha_{OR}=0.4t$, $t_{mx}=t_{my}=0$, $\mu=-0.4t$ and $\gamma=0.1$.  For $\alpha_{OR}\neq 0$ and  $t_{mx(my)}=0$, $f_{2,2_x}$ and $f_{2,2_y}$ are real for any $\bk$, while $f_{2,2_z}$ has a non-zero imaginary part in k-space, arising from the non-zero inter-orbital pairing amplitude $f_{ab}(\bk)$.}
\label{fig:4}
\end{center}
\end{figure*}

Let us start by considering the case of a ${\mathcal C}_{4v}$-symmetric
two-dimensional acentric superconductor 
with a nonvanishing orbital Rashba coupling ($\alpha_{\mathrm{OR}}$). 
In this case,
we employ the model Hamiltonian in Eq. \ref{ham} assuming that the term $t_{m}$ is 
vanishing. We aim to study 
the structure of the resulting pairing amplitude for the Cooper pairs with net components of the orbital angular moment and maximal projection along the various crystallographic directions.
The starting point is to self consistently determine the order parameters for a given amplitude of the pairing interaction. As mentioned above, we
have that the order parameters $\Delta_{\alpha\alpha}$ are non-vanishing while $\Delta_{\alpha\beta}$, with $\alpha\neq\beta$, are identically zero (see Fig. \ref{fig:3}a). The latter is due to the symmetry enforced by the vertical mirror and rotation transformations. When considering the $k$-resolved pairing amplitudes in momentum space, however, we find that they can be non-vanishing. Even more importantly, they are non-trivial. In
particular, for one of the inter-orbital channels we have that $f_{ab}(\bk)\neq 0$.
The occurrence of non-vanishing inter-orbital pairing components has interesting consequences on the structure of the quintet states in the momentum space. Let us recall that the structure of the pairing amplitude for Cooper pairs having $J=2$ and maximal orbital polarization ($J_z=2,J_x=2,J_y=2$) has an intrinsic complex phase in momentum space (see Eqs. \eqref{eq6}-\eqref{eq11}). Indeed, the real part is generally provided by the equal-orbital components while the imaginary part is due to the off diagonal pairing amplitudes in the orbital space. The analysis of the various quintet states with maximal orbital polarization 
for a representative value of the orbital Rashba coupling is displayed
in Fig. \ref{fig:4}. We find that the pairing amplitudes for the quintet state with planar ($x,y$) projections (Figs. \ref{fig:4}a,b) have only real components and exhibit an amplitude modulation in momentum space especially when moving along the high symmetry lines from $\Gamma=(0,0)$ to the $X=(0,\pi)$ or $Y=(0,\pi)$ points. On the other hand, the quintet state with out-of-plane orbital polarization ($f_{2,2_z}$) has a nontrivial phase texture in the Brillouin zone (Fig. \ref{fig:4}c). We find that the vector field constructed from the real and imaginary components of the $f_{2,2_z}$ pairing amplitude can form a distinct pattern in momentum space. A peculiar phase winding nucleates close to the points $P=(\pm\pi/2,\pm \pi/2)$ and $Q=(\pm\pi/2,\mp \pi/2)$ along the diagonals of the Brillouin zone in proximity of the avoiding crossings of the electronic bands. This profile is due to the orbital mixing that is enhanced in that particular region of the Brillouin zone and the amplitude of the unequal-orbital pairing amplitudes is boosted. 
A close inspection of the phase pattern  indicates that the texture is substantially guided by a $\pi$-phase shift of the vector fields when moving along the lines parallel to the $k_x$ and $k_y$ orientations across the $P$ and $Q$ points. Close to the $P$ and $Q$ points there is a nontrivial phase rotation, with a $\pi/2$ offset, in the connecting phase domains when moving from the boundary at $k_x=\pi,k_y=\pi/2$ to the boundary at $k_x=\pi/2,k_y=\pi$.   

\begin{figure*}[htb]
\begin{center}
\includegraphics[width=0.98\textwidth]{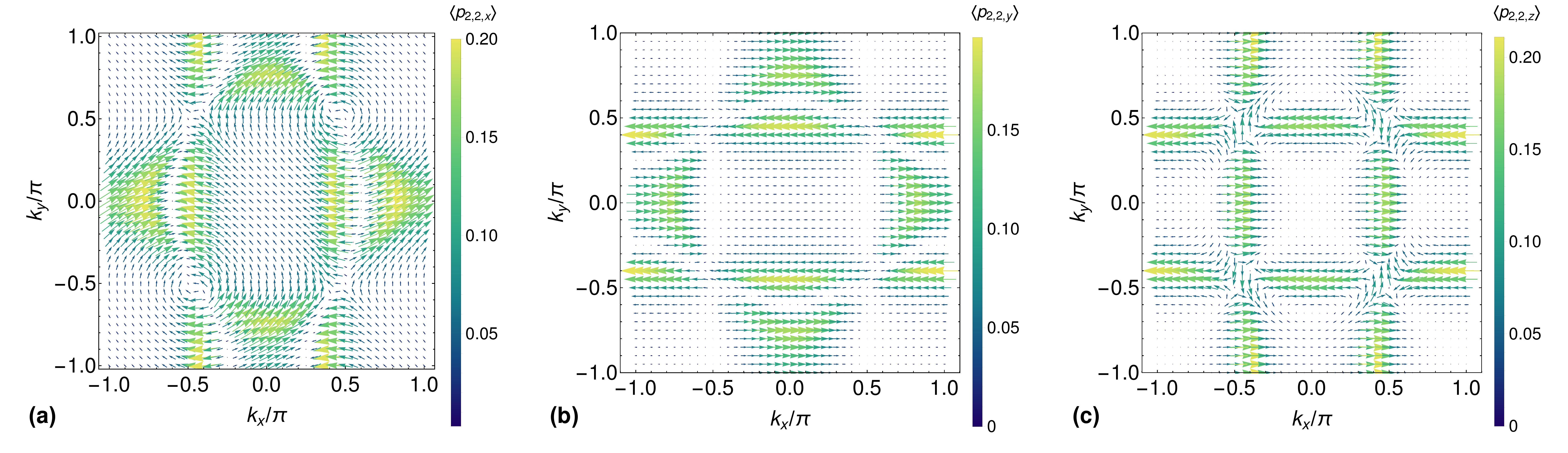} 
\protect\caption{(a)-(c) Vector fields of the pairing amplitudes $f_{2,2_\alpha}=\langle p_{2,2_\alpha}\rangle$ with ($\alpha=x,y,z$). The vectors components are given by the real and imaginary part of the pairing amplitude $\langle p_{2,2,\alpha}\rangle$, i.e. $(\rm{Re}\langle p_{2,2,\alpha}\rangle,\rm{Im}\langle p_{2,2,\alpha}\rangle)$. Here the parameters  $\alpha_{OR}=0.4t$, $t_{mx}=0.2t$, $t_{my}=0$, $\mu=-0.4t$ and $\gamma=0.1$.   For $t_{mx}\neq 0$, also $f_{2,2_x}$ and $f_{2,2_y}$ have a non-zero imaginary part in k-space (with $\text{Im}(f_{2,2_y})$ being much smaller in amplitude than $\text{Re}(f_{2,2_y})$).}
\label{fig:5}
\end{center}
\end{figure*}

\begin{figure*}[htb]
\begin{center}
\includegraphics[width=0.98\textwidth]{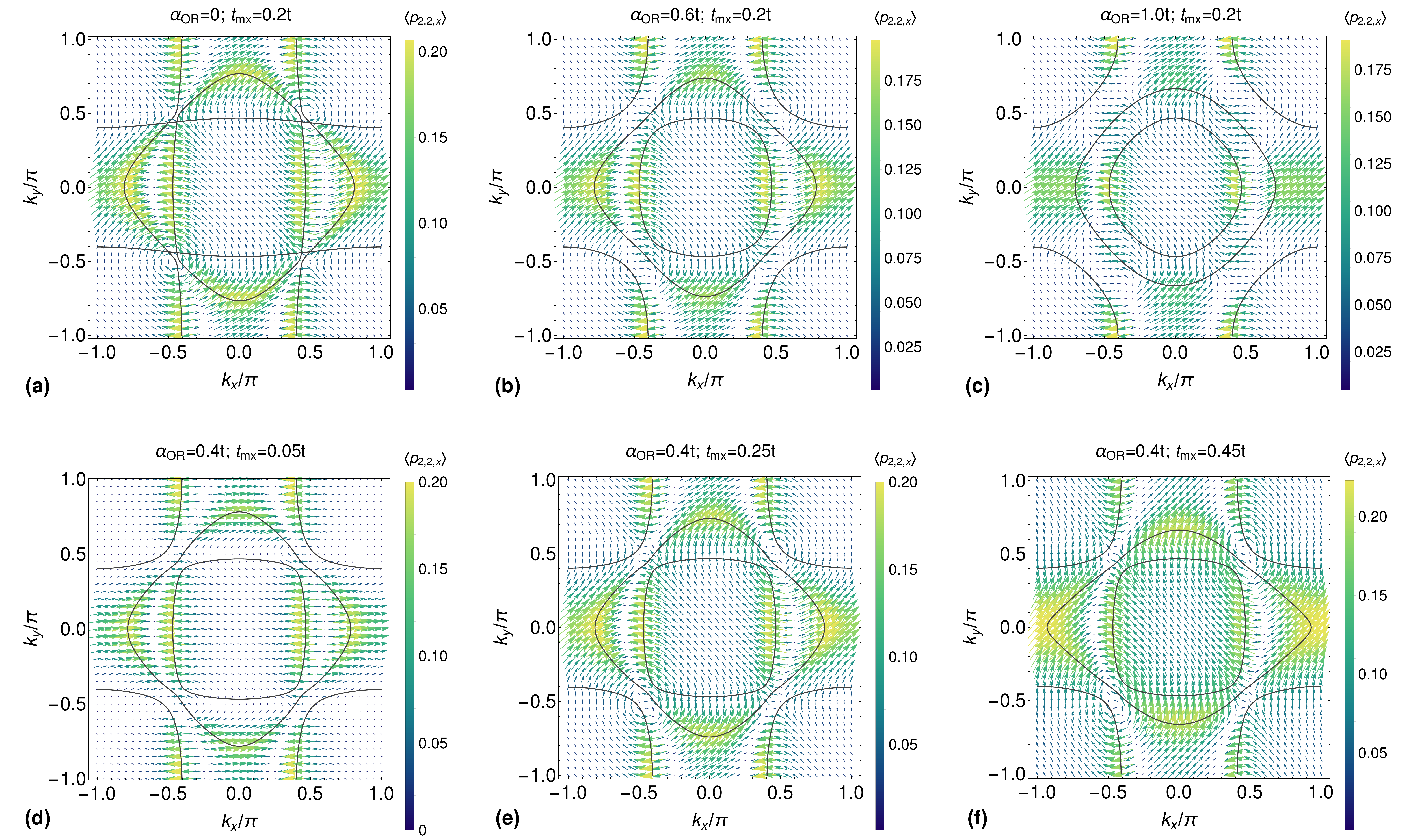} 
\protect\caption{
Vector field pattern of the pairing amplitude $f_{2,2_x}=\langle p_{2,2_x}\rangle$. The vector's components are given by $(\rm{Re}(f_{2,2_x}),\rm{Im}(f_{2,2_x}))$. In (a)-(c) we show the evolution of the orbital quintet texture in momentum space as a function of the orbital Rashba coupling $\alpha_{OR}$, at a given value of the parameter $t_{mx}$, while in (d)-(f) the orbital Rashba coupling is fixed and we show the evolution as a function of $t_{mx}$. 
The black lines show the Fermi lines of the normal state.
}
\label{fig:6}
\end{center}
\end{figure*}

We then switch on the term that allows us to break one vertical mirror symmetry as well as the $C_4$ rotation, while leaving the residual vertical mirror symmetry behind.
Such crystalline symmetry lowering is implemented by the presence of the $t_{mx}$ coupling in our model. As mentioned above, the $t_{mx}$ term preserves the $\mathcal{M}_x$ mirror symmetry while it breaks both $\mathcal{M}_y$ and $C_4$ 
For this physical situation, we have that the inter-orbital order parameter  $\Delta_{bc}$ is also nonvanishing  (see Fig. \ref{fig:3}b-c). Moreover, when considering the pairing amplitude in momentum space, we find that they are nontrivial and nonvanishing in all intra- and inter-orbital channels. Hence,
the further reduction of the point-group symmetry from $\mathcal{C}_{4v}$ to $\mathcal{C}_{s}$ enriches the structure of the pairing amplitude of quintet states in the momentum space. 
Indeed, for the quintet state with $x$-orbital component, there is a line of zeros for the real part of the pairing amplitude that separates two half-vortices with opposite winding and the cores centered close to the points $P$ and $Q$ in the Brillouin zone (see Fig. \ref{fig:5}a). 
On the other hand, the pattern of $y$- and $z$- orbitally polarized quintet states is not much affected by the further symmetry breaking due to the inclusion of the $t_{mx}$ term (Fig. \ref{fig:5}b,c).  
It is interesting to observe how the vortex-like structures evolve as a function of the strength of the couplings that control the lowering of the point-group symmetry when moving from $C_4$ to $\mathcal{C}_{4v}$ and then to $\mathcal{C}_{s}$.
In Fig. \ref{fig:6} we show the evolution of the quintet pairing amplitude texture associated to the in-plane ($x$) orbital moment polarization as a function of the orbital Rashba coupling at a given value of $t_{mx}$. We observe that the vortex-like structure of the phase texture along the diagonals of the Brillouin zone is deformed by the modification of the line of vanishing pairing amplitudes as $\alpha_{OR}$ increases (Fig. \ref{fig:6}a,b,c). A flipping of the phase larger than $\pi/2$ is observed when moving along the $\Gamma-X$ directions in the Brillouin zone with a $\pi$ phase transition at a given threshold of the momentum. Such type of texture phase change is not observed along the $\Gamma-Y$ direction where the orbital texture exhibits a phase slip of about $\pi/2$ in proximity of the Fermi contour.
Another consequence of the increase of the orbital Rashba coupling is that it creates an accumulation of the orbital polarization in a smaller portion of the Brillouin zone nearby the Fermi lines. 

When considering the impact
of the $t_{mx}$ term we clearly observe how the vorticity around the $P$ and $Q$ points of the Brillouin zone sets out by increasing its amplitude from small to large values (Fig. \ref{fig:6}d,e,f). We also find a slight deformation of the phase winding around the $P$ and $Q$ points of the Brillouin zone. The amplitude of the $x$-polarized quintet is amplified in proximity of the zone boundary at $k_y\sim 0$ with a $\pi$-phase jump occurring when moving from the $\Gamma$ to the boundary of the Brillouin zone.

\section{Josephson effects with orbitally polarized Cooper pairs}

\begin{figure}[t!]
\begin{center}
\includegraphics[width=0.98\columnwidth]{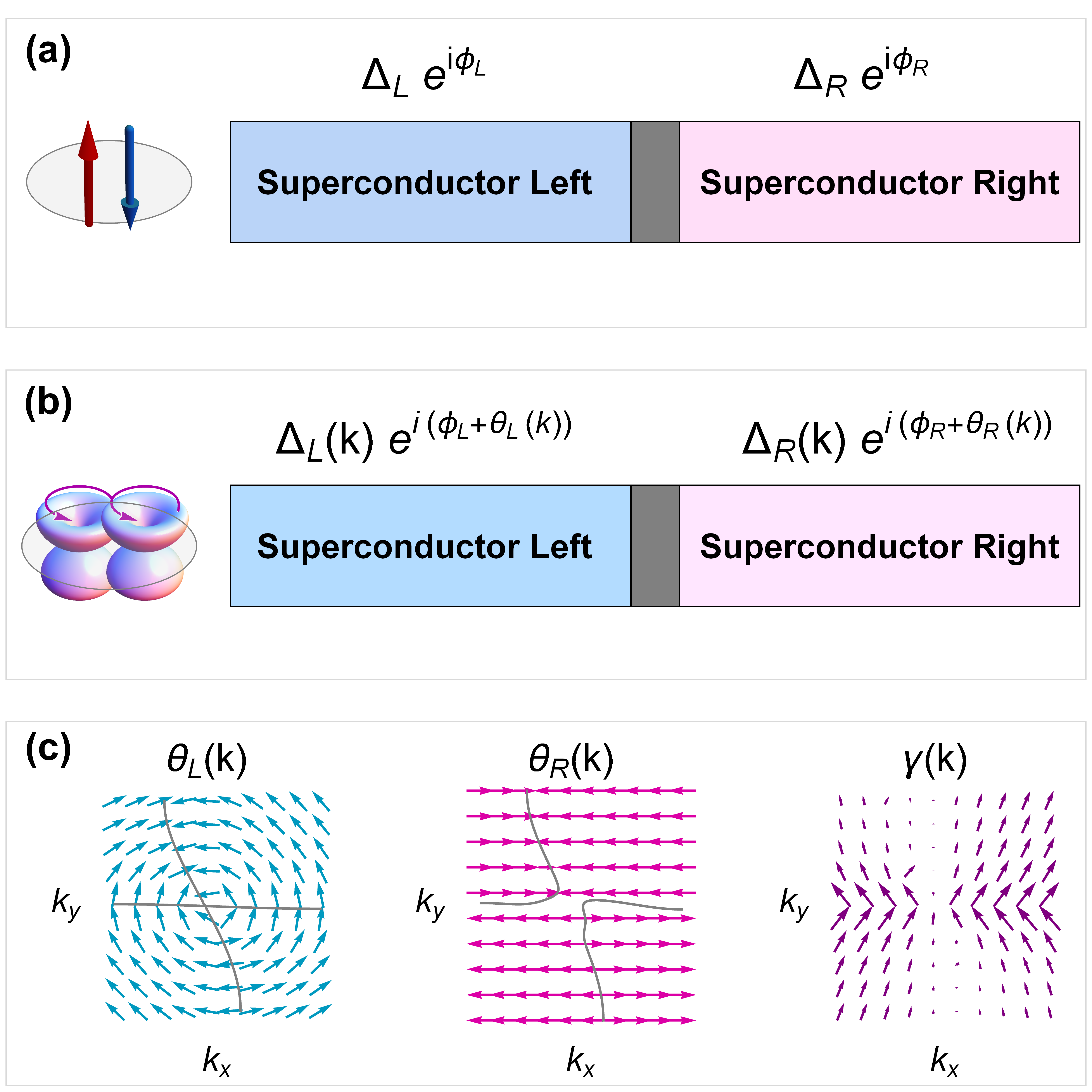} 
\protect\caption{ (a) Scheme of a conventional Josephson junction with a phase bias $\phi_L-\phi_R$ across the interface. 
(b) Sketch of the Josephson junction with high orbital moment Cooper pairs and made with two superconductors with nonequivalent strength of the
crystalline symmetry breaking. In this case there will be different phase textures $\theta(\bk)$ when comparing the left and right sides of the superconducting junction. (c) Plots of $\theta_L(\bk)$, $\theta_R(\bk)$  and $\gamma(\bk)=\theta_{L}(\bk)-\theta_{R}(\bk)$ in a portion of the Brillouin zone and for a representative case of the electronic parameters. 
Tunneling of Cooper pairs at a given momentum $\bk$ can acquire an intrinsic nontrivial phase offset $\gamma(\bk)$. The gray lines in the plots of $\theta_{L(R)}(\bk)$ are the nodal lines of the real part of $f_{2,2_x}(\bk)$. Nearby these nodal lines $\pi$-phase walls develop, assuming that the amplitude of $\rm{Im}(f_{2,2_x})$ is small.}
\label{fig:7}
\end{center}
\end{figure}

Let us now consider the consequences of the achieved orbital textures in a Josephson junction  obtained by interfacing two superconductors with nonequivalent crystalline symmetries or strength of the symmetry breaking coupling.  
In the presence of conventional Cooper pairing the current-phase relation (CPR) of a superconductor-insulator-superconductor junction is given by $I_J=I_c \sin(\phi)$ \cite{Jos62}, with $I_c$ being the critical current and $\phi=(\phi_L-\phi_R)$ the phase difference between the left and right lead of the junction (Fig. \ref{fig:7}a).
Hallmarks of the CPR in conventional Josephson junction are provided by a vanishing supercurrent and non-degenerate minimum of the Josephson energy at $\phi=0$. Deviations from standard CPR are commonly obtained in the presence of sources of time-reversal symmetry breaking. They can be marked by a Josephson energy offset of a fractional flux quantum, leading to the so called $\varphi_0$-junction \cite{Buzding2008,Tanaka2009,Tanaka2014,Golubov2004,Tanaka2000,Asano2003,Grein2009,Eschrig2008,Silaev2017,Konschelle2019,Alidoust2021} or by half-integer flux quantum offset, leading to the so-called $\pi$-junction, due to magnetic exchange \cite{Golubov2004,Buzdin2005,Bergeret2005} or Rashba spin-orbit fields \cite{Sothmann2015,Yang2008}.
Apart from anomalous phase shifts, the energy of the Josephson junction can keep the phase-inversion symmetry but with a minimum at values of the Josephson phase which is different from the high symmetry values, 0 or $\pi$, setting out a $\varphi$-Josephson junction.
From a general point of view, a $\varphi$-junction is achieved by: i) combination of 0- and $\pi$- Josephson couplings {{\cite{Goldobin2007,Goldobin2011,Yerin2014,Sickinger2012,Guarcello2022}}} ; ii) by employing suitable parameters range and junction geometries \cite{TanakaKashi96,TanakaKashi97,Josephson2,KashiwayaTanaka2000RepProgPhys}; or iii) by means of higher harmonics \cite{Goldobin2007,fukaya22}. The use of inversion symmetry breaking in topological superconductors or at the interface of conventional superconductors can also lead to $\pi$ or $\varphi$-Josephson junctions.
In other words,
the generation of nontrivial Josephson phase couplings typically requires 
either an unconventional type of superconductor or the use of magnetism in superconducting heterostructures.

Our study allows to individuate
another path to achieve anomalous Josephson effects: it 
exploits orbital quantum resources of the Cooper pairs. Indeed, we demonstrate that multiorbital spin-singlet superconductors can be marked by an intrinsic $k$-dependent phase with a nontrivial texture in momentum space when considering high-orbital moment Cooper pairs. Since the superconducting phase pattern and orbital polarization of the Cooper pairs depend on the strength and character of the point-group symmetry breaking couplings, one can use these quantum resources to set out non standard Josephson effects (Fig.\ref{fig:7}b). One path to achieve an anomalous Josephson coupling is to design junctions that have different degree of crystalline symmetry breaking due to intrinsic or extrinsic sources. For instance, assuming that the couplings related to the point-group crystalline symmetry breaking have unequal amplitude on the two sides of the junction, we have that tunneling of Cooper pairs can experience a Josephson phase offset that arises from the intrinsic orbital phase texture. The crystalline junction asymmetry implies that, for a given momentum $\bk$ in the Brillouin zone, the phase $\theta(\bk)$ associated with the quintet pairing does not have the same value on the left and right side of the junction (Fig.\ref{fig:7}c). Hence, one has that the difference, $\gamma(\bk)=\theta_{L}(\bk)-\theta_{R}(\bk)$, would be generally 
different from zero in the whole Brillouin zone apart from sparse
accidental points. Then, if we restrict 
ourselves to almost $k$-dependent
tunneling processes which preserve the orbital moment polarization across the interface, there will be an overall phase shift by summing up all the phase mismatches close to the Fermi contour. The acquired total phase, $\gamma_i$ (i=x,y,z), depends on the orientation of the quintet pairs, and will lead to a Josephson current of the type $J_{2,2_i} \sim J_i \sin(\phi+\gamma_i)$, where $\gamma=\int_{\Gamma} \gamma_i(k) dk$, $J_i$ is related to the tunneling processes, and the integration domain $\Gamma_k$ indicates the set of momenta in the Brillouin zone corresponding to the regions of intersections of the Fermi contours of the superconductors in the left and right side of the junction. We point out that in time-reversal symmetric conditions the component with Cooper pairs having opposite orbital polarization will lead to a Josephson current with opposite phase, i.e. $J_{2,-2_i} \sim J_i \sin(\phi-\gamma_i)$. Hence, we can deduce that the superposition of the Josephson currents with opposite orbital polarization would lead to a changeover from $0$ to $\pi$ of the Josephson coupling if the accumulated phase $\gamma_i$ is larger the $\pi/2$. Another general consequence about the Josephson effect is that the interference of the orbital dependent channels can lead to dominant high harmonics as the occurrence of $\pi$ Josephson phases tends to generate a cancellation of the first harmonic contribution. 

\begin{figure}[hbt]
\begin{center}
\includegraphics[width=0.98\columnwidth]{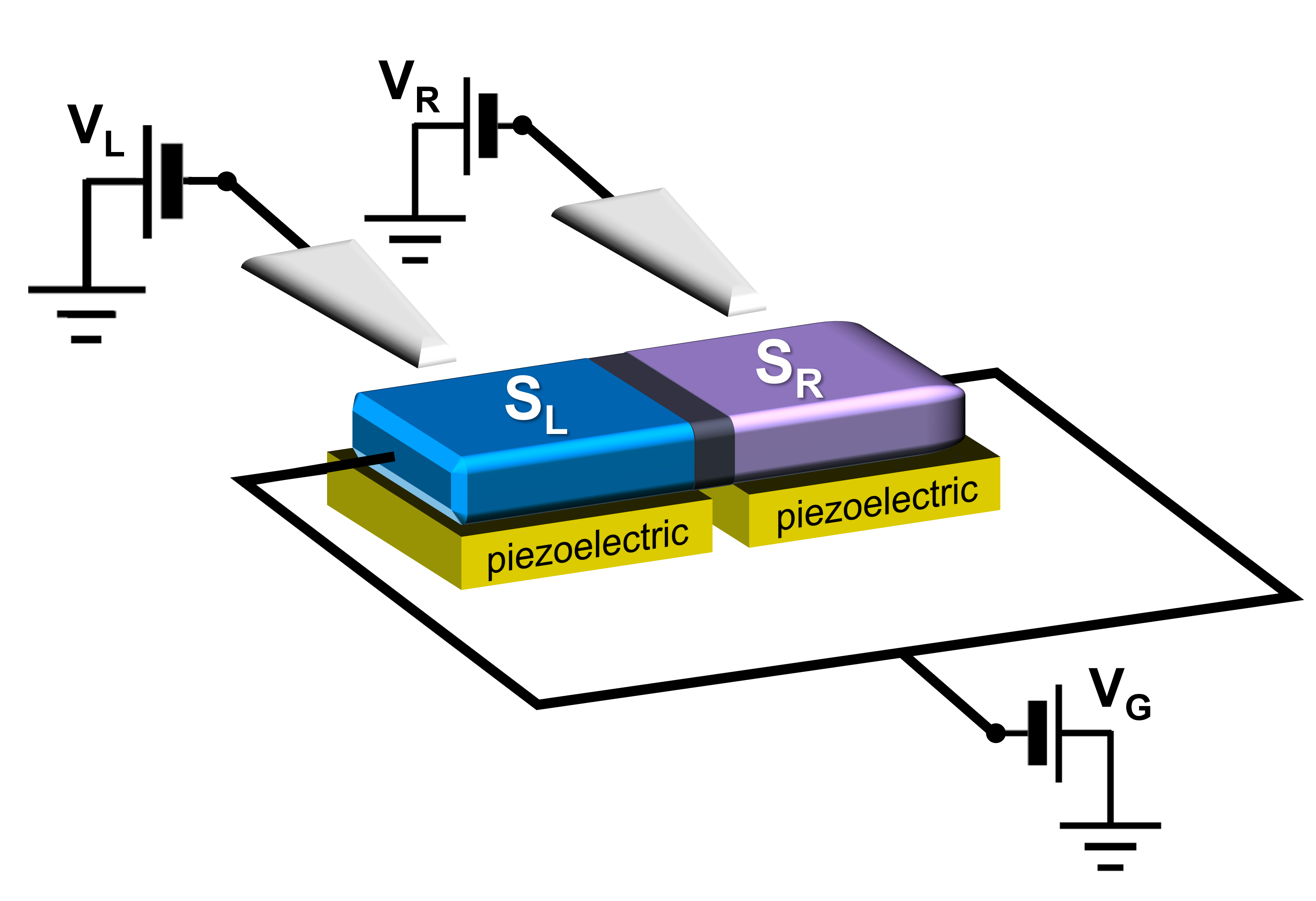} 
\protect\caption{ 
Scheme of a setup for probing high orbital moment Cooper pairs by means of gate and strain driven Josephson effects. The Josepshon device is made of tunnel coupled superconductors ($S_L$ and $S_R$) which can be subjected to an electrostatic potential (via gates) and/or strain field (trough interfaced piezoelectric). The superconducting junction is inserted into a loop circuit with a SQUID type geometry to measure the current phase relation of the junction. The gate and strain fields can be effectively employed to control the strength of the crystalline symmetry breaking potentials as related to $\alpha_{OR}$ and $t_{m}$. The measurement of the current phase relation as a function of the external drives would provide indications of anomalous Josephson effects mediated by the presence of high-orbital moment Cooper pairs due to the nontrivial phase textures.
}
\label{fig:8}
\end{center}
\end{figure}

A setup which can be employed to probe anomalous Josephson effects as due to the presence of high-orbital moment Cooper pairs is schematically shown in Fig. \ref{fig:8}. By means of gate and strain fields one can suitably control the degree of crystalline symmetry breaking in the superconducting leads forming the Josephson junction. Hence, if the junction is inserted in a SQUID loop, one can employ conventional approaches to measure the current phase relation. As discussed before, we foresee an anomalous current phase relation with reversal of the sign and suppression of amplitude of the supercurrent.

\section{Conclusions}

We have demonstrated that multiorbital 2D spin-singlet superconductors in the presence of unusually low degree of point-group crystal symmetry host Cooper pairs with net components of the orbital angular momentum. Although we have analyzed the specific case of the changeover from  $\mathcal{C}_{4v}$ to $C_{s}$ point-group crystal symmetries, the achieved outcomes can be generalized to materials with other point-group symmetries. 
We find that the lack of inversion, mirror and rotation symmetries drives the formation of quintet pairs with a net component of the orbital polarization and a nontrivial intrinsic phase texture in momentum space. This phase texture of the quintet states can be exploited in Josephson junction to design $0$ to $\pi$ phase transitions in the Josephson coupling. 
If the crystalline environment does not provide symmetry breaking interactions that are sufficiently strong, we expect that the application of external electric fields or strains can be the most impactful means
to yield pairs with quintet orbital polarization.
Hence, electric fields and applied strains are suitable knobs to tune the Josephson
coupling, e.g. by inducing the 0-$\pi$ transitions or by generating spontaneous 
orbital polarized supercurrents in the junction. The presence of an intrinsic orbital phase texture in the momentum space can also drive a suppression of the supercurrent due to 
orbital phase interference.
Recently, gate induced supercurrent suppression in ultrathin metallic superconductors has been observed \cite{DeSimoni2018,Golokolenov2021,Alegria2021,Elalaily2021,Basset2021,Ritter2022} for a large variety of materials and setup configurations \cite{Ruf2023}, including free standing nanowires \cite{Rocci2020} and liquid ionic gating \cite{Paolucci2021}. While various dynamical mechanisms have been proposed to account for the observed phenomenology, the overall anomalies of the uncovered effect indicate a nontrivial interplay of superconductivity, point group symmetry breaking, and electric or strain fields.

Apart from elemental superconductors in the presence of applied gate or strains, other low-dimensional materials can be suitable candidate to host a superconducting phase with high orbital angular momentum. For instance, 2D noncentrosymmetric oxides heterostructures, such as LaAlO$_3$/SrTiO$_3$~\cite{Ohtomo-2004, Reyren-2007,Caviglia-2008} are promising candidates.
In particular, oxides interfaces grown along the
[111] crystallographic orientation 
are naturally equipped with orbital degrees of freedom and the required low crystalline symmetry \cite{Mercaldo2023,Lesne2023}. Such matching applies to oxides interface with SrTiO$_3$~\cite{mon19,kha19}, KTaO$_3$~\cite{liu21}, and SrVO$_3$-based heterostructures that host two-dimensional $d$ electron systems with ($d_{xy},d_{xz},d_{yz})$ orbitals and superconductivity.

The ability 
to induce 0-$\pi$ transitions in Josephson device by means of gate or strains that tune the strength of the crystalline symmetry breaking in the superconducting leads can have interesting perspectives for the design of innovative 0-$\pi$ qubit. It is known that superconducting circuits with quantum qubits based on the Josephson effect represent a promising platform for quantum computing. One of major challenge in quantum information processing is to keep the coherence of quantum superposition over long times by for instance optimally reducing the noise effects.   
Noise-protected qubits should have a high degree of protection both against energy relaxation and pure dephasing. 
The achievement of such protection, however, can result in specific constraints on qubits that are equipped with a single degree of freedom.
Indeed, differently from single-degree qubits, as for the case of transmon, superconducting circuits with more
degrees of freedom can be simultaneously
protected against energy relaxation and pure dephasing. One promising candidate for such a protected superconducting qubit is the $0-\pi$ circuit \cite{Brooks_piqubit_2013} which has been recently realized in an experimentally achievable regime \cite{Gyenis_piqubit_2021}. Along this line, proposals to implement the 0-$\pi$ qubit have been considering the spin degrees of freedom of the Cooper pairs in super-conductor-ferromagnet or in superconductor-semiconductor heterostructures \cite{Cai2023,Guo_piqubit2022}. 
Taking the results of our study as a bedrock,
we envision the possibility to design 0-$\pi$ qubit by exploiting the orbital degrees of freedom of spin-singlet Cooper pairs. 
The fact that the $\pi$ Josephson phase can be controlled through a modification of orbital dependent 
coupling additionally indicates the possibility of having electrically driven 0-$\pi$ qubits.

\appendix
\section{Addition of $L=1$ angular momentum and pairing operators}

\begin{figure}[htb]
\begin{center}
\includegraphics[width=0.45\textwidth]{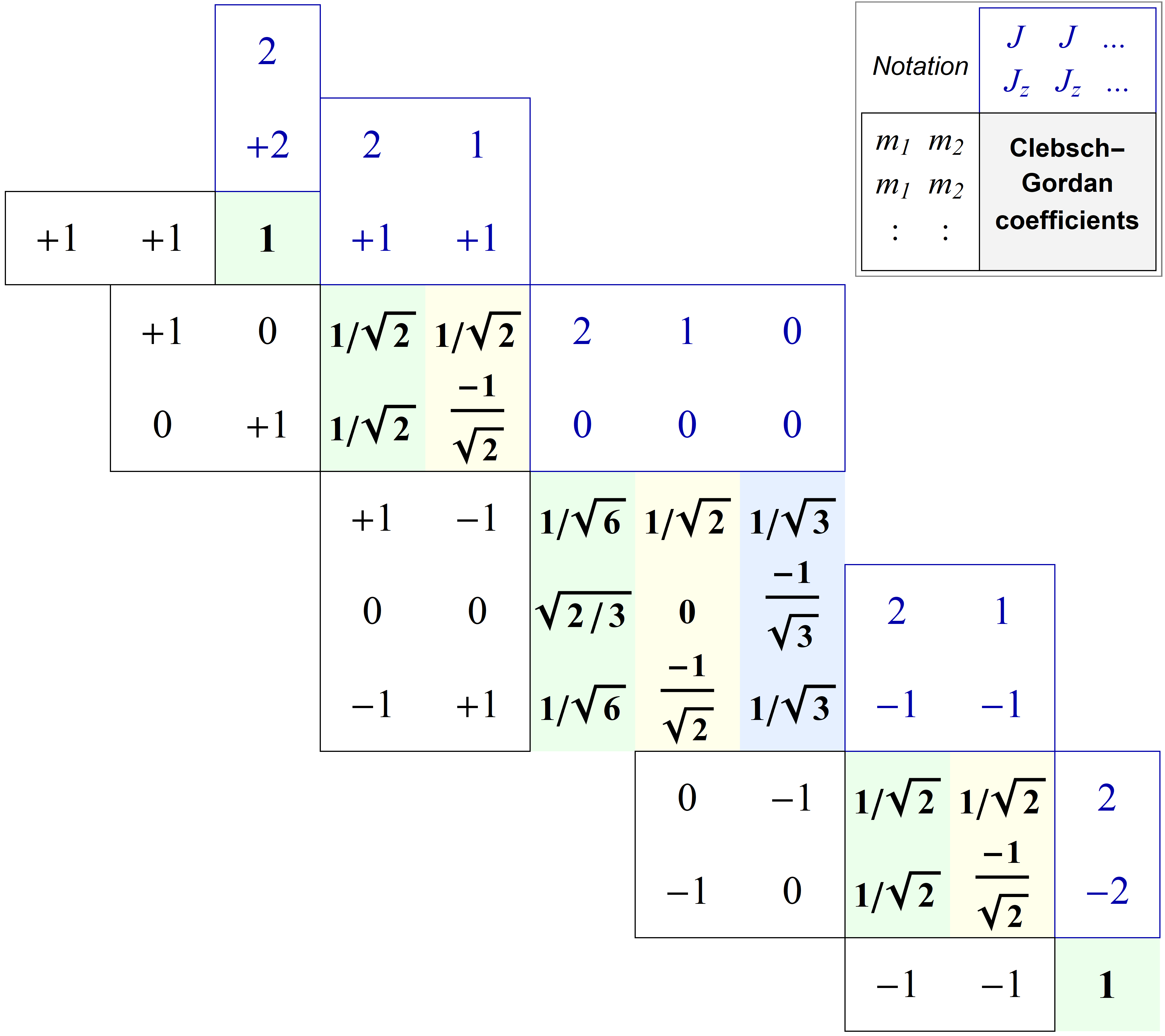} 
\protect\caption{Clebsch-Gordan coefficients tables for $L_1=L_2=1$ (n.b. $m_1+m_2=J_z$). The coefficients for the quintet $J=2$ are highlighted in green, the triplet $J=1$ in yellow and those for the state $J=0$ in light-blue. For example: $|1,1\rangle= (1/\sqrt{2})(|1,+1;1,0\rangle - |1,0;1,+1\rangle$.}
\label{fig:A1}
\end{center}
\end{figure}

In this Appendix we provide details about  the form of the pairing operators as deduced by the addition of $L=1$ angular momentum.
The single electron description of the model Hamiltonian is based on three orbitals spanning an $\hat{L}=1$ angular momentum subspace. We use  $|1,0_j \rangle,\, (j=x,y,z)$ to denote a single particle Wannier state with zero projections of the angular momentum $\hat{L}_j$.
In an analogous way, one can also construct the electronic state with $\pm 1$ quantum numbers for the corresponding components of the angular momentum.
For instance for $L_z$ one has that $|1,\pm1_z\rangle = \frac{1}{\sqrt{2}}(|1,0_y\rangle \mp i |1,0_x\rangle)$, with $\hat{L}_z|1,\pm1_z\rangle = (\pm 1) |1,\pm1_z\rangle$  (in unit of $\hbar$).
These states can be also expressed in terms of occupation number by introducing the creation 
operators acting on the vacuum state: 
\begin{eqnarray}
\label{eq:rel1}
    \begin{array}{cc}
d_{+1_{z},\sigma}^\dagger=\frac{1}{\sqrt{2}}(c_{b\sigma}^\dagger - i\, c_{a\sigma}^\dagger)   \,,  &  \quad d_{-1_{z},\sigma}^\dagger=\frac{1}{\sqrt{2}}(c_{b\sigma}^\dagger + i\, c_{a\sigma}^\dagger) \,, 
    \end{array}
\end{eqnarray}
with $|1,0_z\rangle$ being associated with  $c^\dagger_{c,\sigma}$.
Similar relations hold for the other two components of the orbital angular momentum, i.e. $\hat{L}_x$ and $\hat{L}_y$.  Specifically, $|1,\pm1_x\rangle = \frac{1}{\sqrt{2}}(|1,0_z\rangle \pm i |1,0_y\rangle)$ and $|1,\pm1_y\rangle = \frac{1}{\sqrt{2}}(|1,0_z\rangle \mp i |1,0_x\rangle)$, and hence $d^\dagger_{\pm_x,\sigma}=\frac{1}{\sqrt{2}}(c^\dagger_{c\sigma}\pm i\, c^\dagger_{b\sigma})$ and $d^\dagger_{\pm_y,\sigma}=\frac{1}{\sqrt{2}}(c^\dagger_{c\sigma}\mp i\, c^\dagger_{a\sigma})$.
Following the standard procedure for the addition of angular momentum, we can write the state $|J,J_z\rangle$ in terms of the $|L_1,m_1;L_2,m_2\rangle$ ones by using the Clebsch-Gordan coefficients 
(see Fig. \ref{fig:A1})

Considering the relation between the single particle states $|1,m\rangle \; (m=\pm1,0)$ and the operators $d^\dagger_{\pm_z,\sigma}$ (for $m=\pm1)$ and $c^\dagger_{c,\sigma}$ (for $m=0$),
one can construct the pair operators by combining them.
Indeed, the pair operators $p_{J,J_\alpha}$ for evaluating the superconducting order parameter in Sec. \ref{sec:s3} are written in terms of annihilation operators, thus $p_{2,2_z}=d_{+1_z,\up}\,d_{+1_z,\dw}$ which can be in turn expressed in term of the basis operators $c_{\alpha\sigma}$ through Eq.\eqref{eq:rel1}, resulting in the relation reported in Eq. \eqref{eq:p22z}, as well as $p_{1,1_z}=\frac{1}{\sqrt{2}}(d_{+1_z,\up} c_{c_\dw} - c_{c\up} d_{+1_z,\dw}$ can be rewritten as in Eq.\eqref{eq11}.
Using the 
expressions for the $|J,J_z\rangle$ states, which can be obtained using the Clebsch-Gordan coefficients (Fig.\ref{fig:A1}), 
is then possible to deduce the form of all the $p_{J,J_z}$ operators and a similar construction holds for  
$p_{J,J_x}$ and $p_{J,J_y}$.

\medskip
\noindent 
{\bf{Acknowledgments}}
\\
M.C. and M.T.M. acknowledge support from the EU’s Horizon 2020 research and innovation program under Grant Agreement No. 964398 (SUPERGATE). M.C. acknowledges financial support from PNRR MUR project PE0000023-NQSTI. 


\end{document}